\documentclass{aa}

\usepackage{graphicx}

\begin{document}

\title{The unidentified TeV source (TeV~J2032+4130) and surrounding field: Final HEGRA IACT-System results.}

\titlerunning{TeV~J2032+4130}

\author{
F.~Aharonian\inst{1},
A.~Akhperjanian\inst{7},
M.~Beilicke\inst{4},
K.~Bernl\"ohr\inst{1},
H.-G.~B\"orst\inst{5},
H.~Bojahr\inst{6},
O.~Bolz\inst{1},
T.~Coarasa\inst{2},
J.~Contreras\inst{3},
J.~Cortina\inst{10},
S.~Denninghoff\inst{2},
V.~Fonseca\inst{3},
M.~Girma\inst{1}
N.~G\"otting\inst{4},
G.~Heinzelmann\inst{4},
G.~Hermann\inst{1},
A.~Heusler\inst{1},
W.~Hofmann\inst{1},
D.~Horns\inst{1},
I.~Jung\inst{1,9},
R.~Kankanyan\inst{1},
M.~Kestel\inst{2},
A.~Kohnle\inst{1},
A.~Konopelko\inst{1,14},
D.~Kranich\inst{2},
H.~Lampeitl\inst{4},
M.~Lopez\inst{3},
E.~Lorenz\inst{2},
F.~Lucarelli\inst{3},
O.~Mang\inst{5},
D.~Mazin\inst{2},
H.~Meyer\inst{6},
R.~Mirzoyan\inst{2},
A.~Moralejo\inst{3},
E.~O\~na-Wilhelmi\inst{3},
M.~Panter\inst{1},
A.~Plyasheshnikov\inst{1,8},
G.~P\"uhlhofer\inst{11},
R.~de los Reyes\inst{3},
W.~Rhode\inst{6},
J.~Ripken\inst{4},
G.~P.~Rowell\inst{1},
V.~Sahakian\inst{7},
M.~Samorski\inst{5},
M.~Schilling\inst{5},
M.~Siems\inst{5},
D.~Sobzynska\inst{2,12},
W.~Stamm\inst{5},
M.~Tluczykont\inst{4,13},
V.~Vitale\inst{2},
H.J.~V\"olk\inst{1},
C.~A.~Wiedner\inst{1},
W.~Wittek\inst{2}}

\institute{Max-Planck-Institut f\"ur Kernphysik, Postfach 103980, D-69029 Heidelberg, Germany
\and Max-Planck-Institut f\"ur Physik, F\"ohringer Ring 6, D-80805 M\"unchen, Germany
\and Universidad Complutense, Facultad de Ciencias F\'{\i}sicas, Ciudad Universitaria, E-28040 Madrid, Spain
\and Universit\"at Hamburg, Institut f\"ur Experimentalphysik, Luruper Chaussee 149, D-22761 Hamburg, Germany
\and Universit\"at Kiel, Inst. f. Experimentelle und Angewandte Physik, Leibnizstr. 15-19, D-24118 Kiel, Germany
\and Universit\"at Wuppertal, Fachbereich Physik, Gau{\ss}str.20, D-42097 Wuppertal, Germany
\and Yerevan Physics Institute, Alikhanian Br. 2, 375036 Yerevan, Armenia
\and Altai State University, Dimitrov Street 66, 656099 Barnaul, Russia
\and Now at Washington University, St. Louis MO 63130, USA
\and Now at Max-Planck-Institut f\"ur Physik, F\"ohringer Ring 6, D-80805 M\"unchen, Germany
\and Now at Landessternwarte Heidelberg, K\"onigstuhl, Heidelberg, Germany
\and Home institute: University Lodz, Poland
\and Now at Laboratoire Leprince-Ringuet, Ecole Polytechnique, Palaiseau, France (IN2P3/CNRS)
\and Now at Humboldt Universit\"at f. Physik, Newtonstr. 15, Berlin, Germany
}
\authorrunning{Aharonian et al.}

\date{Received 29 June 2004 / Accepted 17 September 2004}

\offprints{G.P. Rowell, D. Horns\\
\email{\scriptsize Gavin.Rowell@mpi-hd.mpg.de,Dieter.Horns@mpi-hd.mpg.de}}

\abstract{
The unidentified TeV source in Cygnus is now confirmed by follow-up observations
from 2002 with the HEGRA stereoscopic system of Cherenkov Telescopes.
Using all data (1999 to 2002) we confirm this new source 
as steady in flux over the four years of data taking, extended with radius 6.2$^\prime$ 
($\pm 1.2^\prime_{\rm stat}$ $\pm 0.9^\prime_{\rm sys}$)
and exhibiting a hard spectrum with photon index $-1.9$. It is located in the direction 
of the dense OB stellar association, Cygnus~OB2. Its integral flux above energies $E>1$ TeV amounts
to $\sim$5\% of the Crab assuming a Gaussian profile for the intrinsic source morphology.
There is no obvious counterpart at radio, optical nor X-ray energies, 
leaving TeV~J2032+4130 presently unidentified. Observational parameters of 
this source are updated here and some astrophysical discussion is provided. Also included are upper limits for a number of other
interesting sources in the FoV, including the famous microquasar Cygnus~X-3.
 
}

\maketitle

\section{Introduction}
The reasonably large fields of view (FoV, eg. FWHM $\geq3^\circ$) achieved by ground-based $\gamma$-ray telescopes
permits survey-type observations using a few or even singly pointed observations. Such potential
was realised with the serendipitous discovery of a TeV source in the Cygnus region.
Analysis of archival data ($\sim$121 h from 1999 to 2001) of the HEGRA system of Imaging 
Atmospheric Cherenkov Telescopes (HEGRA IACT-System) 
revealed convincing evidence for an apparently steady, spectrally hard 
($-$1.9 differential photon index)
and possibly extended TeV source (Aharonian et al. \cite{Aharonian:1}). 
Follow-up observations using the HEGRA IACT-System were performed during its final
season of operation (2002) for a total of $\sim$158 hours. Analysis
of these data again reveal the
presence of this source, thus confirming its existence with the HEGRA IACT-System. 
Earlier, observations (1991) with the HEGRA scintillator array (Merck \cite{Merck:1}) revealed a multi-TeV
excess, positionally consistent with Cygnus~X-3. Analysis with improved direction reconstruction 
(Krawczynski \cite{Kraw:1})
revealed this excess ($+4.3\sigma$ pre-trial) as centered roughly $0.5^\circ$ north of Cygnus~X-3. Interestingly, 
the Crimean group 
(using the Cherenkov imaging technique) reported a significant excess ($\sim +6.0\sigma$ pre-trial) $\sim 0.7^\circ$ 
north of Cygnus~X-3 (Neshpor et al. \cite{Neshpor:1}), and recently, the Whipple collaboration also reported an 
excess at the HEGRA position ($+3.3\sigma$) in their archival data of 1989/1990 (Lang et al. \cite{Lang:1}).

We summarise here in some detail our analysis and numerical results for TeV~J2032+4130, and
give a brief astrophysical interpretation. Given the large FoV of observations, deep exposures were also
obtained for a number of other interesting sources. Upper limits for these sources are given.

\section{Data analysis \& results}

The HEGRA IACT-System, de-commissioned in September 2002, consisted of five identical
Cherenkov telescopes (each with 8.5 m$^2$ mirror area) on the Canary Island of La Palma (2200 m a.s.l.).  
Employing the stereoscopic technique, this system  
achieved an angular resolution $<0.1^\circ$ and energy resolution $<15$\% for $\gamma$-rays on an event-by-event basis over
the 0.5 TeV to $>$50 TeV regime. Detailed technical descriptions of the HEGRA IACT-System and performance 
can be found in P\"uhlhofer et al. (\cite{Puhl:1}).

Gamma-ray-like events are preferentially selected 
against those of the dominant isotropic background of cosmic-rays (CR). This is achieved with
cuts on $\theta$, the angular distance between the reconstructed and assumed arrival directions, and
also the {\em mean-scaled-width} $\bar{w}$ (Aharonian et al. \cite{Aharonian:2}), which is a measure of an
image's conformity to a $\gamma$-ray-like shape. 
Event arrival directions are reconstructed using the algorithms described in  
Hofmann et al. (\cite{Hofmann:1}). We {\em a priori} selected algorithm `3' for final analysis, but also
employed the other available algorithms to check consistency of the signal.
So-called {\em tight cuts} were implemented, namely: $\theta<0.12^\circ$, $\bar{w}<1.1$, which are optimal 
for point-like sources in a background-dominated scenario. We also required a minimum $n_{\rm tel}\geq3$ 
telescope images per event for the $\theta$ and $\bar{w}$ calculation (Aharonian et al. \cite{Aharonian:1}).
An estimate of the CR background surviving cuts is made using both the template 
(see Rowell \cite{Rowell:1}) and displaced background models for consistency checks on the source excess.
The displacement background model employs different regions in the FoV for background estimation using 
events from $\bar{w}<1.1$.

In Tables~\ref{tab:numbers} and \ref{tab:numbers2} we summarise details of the TeV source which 
includes the excess significance, the source extension $\sigma_{src}$, both of which are calculated 
at the excess centre of gravity (CoG), and also the energy spectrum and flux. Some results are split 
according to data subsets, 1999 to 2001 (dataset \S1), and 2002 (dataset \S2).

\begin{figure*}[t]
\centering
\includegraphics[width=13cm]{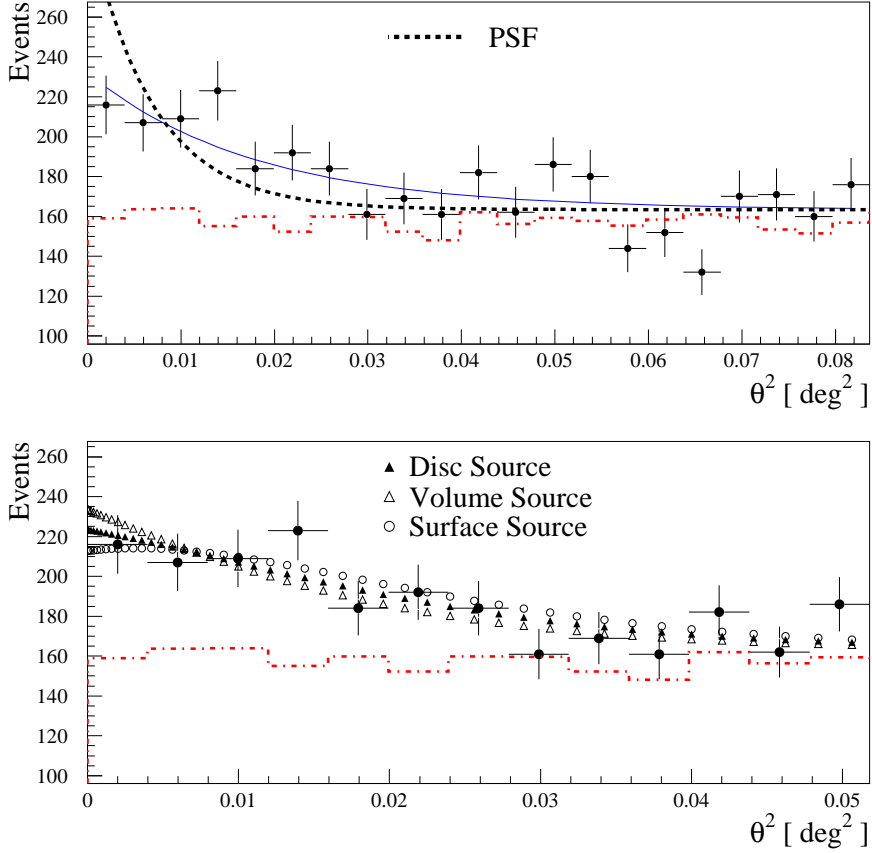}
\caption[]{(Top) Events (solid points with error bars) vs. distance from the CoG squared $\theta^2$ 
           compared with a 
           background estimate from the template model (dashed-dotted line).
           A convolved radial Gaussian fit $F=Ped + P_2 \exp(-\theta^2/(P_1^2+\sigma^2_{\rm pt}))$ is
           indicated by the solid line with $P_1=\sigma_{\rm src}=$0.104$^\circ \
           \pm0.020^\circ$ the intrinsic source size. The PSF width is $\sigma_{\rm pt}$=0.070$^\circ$ 
           (dashed). The pedestal $Ped$ is fitted separately. 
           (Bottom) Comparisons with various intrinsic source morphologies convolved with the PSF (see text).}
\label{fig:extension}
\end{figure*}

The CoG and source extension were estimated by fitting a 2D
Gaussian convolved with the system point spread function (PSF) to a histogram of 
$\gamma$-ray-like ($\bar{w}<1.1$) events binned over a $1^\circ\times 1^\circ$ FoV.
The PSF is estimated from Crab data, giving a value which agrees with
Monto Carlo simulations of a point source (Aharonian et al. \cite{Aharonian:4,Aharonian:5}).
A Gaussian profile suitably describes the excess morphology (discussed later).
Here, we used higher quality events selected according to the estimated error in 
reconstructed direction $\epsilon$ (see eg. Hofmann et al. \cite{Hofmann:2}). This cut on $\epsilon$ reduced somewhat the 
systematic differences in CoG obtained from all three available algorithms 
`1', `2' and `3'.
Simulations and empirical results from other point sources of the HEGRA IACT-System archive have shown that 
consistent sensitivities from all  
algorithms are expected for a minimum number of images per event $n_{\rm tel}\geq3$. 
In fact the signal excess significance from all data using each  
algorithm are consistent to within 1$\sigma$, and the CoGs and source extensions agree to within a 2$\sigma$ level.
Taking the final CoG and source extension from algorithm `3', the 
respective systematic errors were taken using results from the other algorithms.
The excess significance from combined data exceeds 7$\sigma$ (+7.1$\sigma$), and the source appears extended at 
the $>4\sigma$ level when, conservatively, subtracting the systematic error in extension.
The source is termed {\bf TeV~J2032+4130} based on its CoG.
%
%
The post-trial significance is $\sim 6.1\sigma$ if one assumes the trial factors (1100) accrued from the 
discovery dataset \S1 (Aharonian et al. \cite{Aharonian:1}). 
The radial source extension is shown in Fig.~\ref{fig:extension} and compared against a number of possible 
source morphologies. We tested three different types of intrinsic source morphology:
\begin{itemize}
  \item Disc: The source emits at a constant brightness out to a radius 0.13$^\circ$ (with
              zero brightness outside). This profile resembles
              that of a disc.
  \item Volume: The source emits at all positions inside a sphere of radius 0.13$^\circ$. 
  \item Surface: The source emits at all positions within a radial band between 0.117$^\circ$ and 0.13$^\circ$.
\end{itemize}
In each case the volume-integrated radial brightness is calculated after convolution with the instrument
PSF, and compared with the measured radial profile of TeV~J2032+4130.
In all cases similar reduced-$\chi^2 \sim 1.0$ were obtained, thereby not permitting 
discrimination between these source morphologies.

\begin{figure*}
{\begin{center}
 \includegraphics[width=0.70\linewidth]{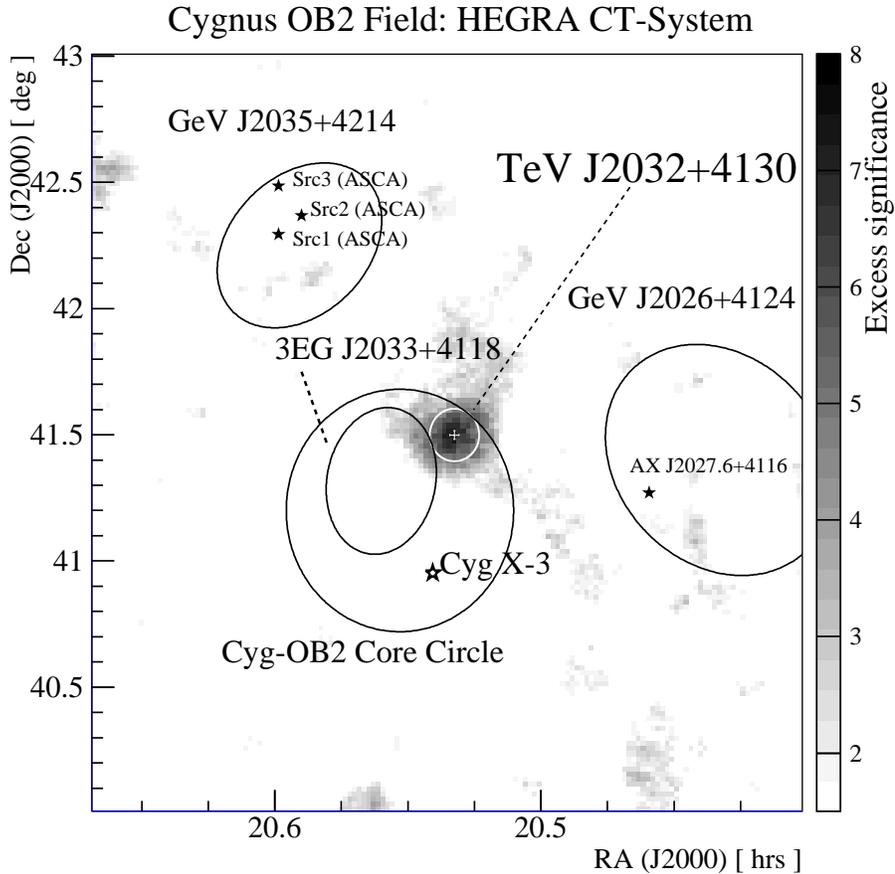}
 \end{center}
}
    \caption{
     Skymap of correlated event excess significance ($\sigma$) from all HEGRA IACT-System data 
     (3.0$^\circ \times 3.0^\circ$ FoV) centred on TeV~J2032+4130. 
     Nearby objects are indicated (EGRET sources with 95\% contours). The TeV source centre of 
     gravity (CoG) with 
     statistical errors, and intrinsic size (std. dev. of a 2D
     Gaussian, $\sigma_{src}$) are indicated by the white cross and white circle, 
     respectively.}
     \label{fig:skymap} 
\end{figure*}

Fig.~\ref{fig:skymap} depicts the 2D {\em skymap} of excess significance over the FoV, using the template
model as a CR background estimate. TeV~J2032+4130 is clearly positioned at the edge of the error 
circle of 3EG~J2033+4118, and within the core circle of the extremely dense OB stellar association Cygnus~OB2
(Kn\"odlseder \cite{Knodlseder:1}).

The energy spectrum determination followed the method described in Aharonian et~al. (\cite{Aharonian:3}) 
using tight cuts (on $\theta$ 
and $\bar{w}$) plus an additional cut on the reconstructed air-shower core distance of the event {\em core}$<$200m.
Reconstruction of event energies employed the more advanced method of Hofmann et al. (\cite{Hofmann:3}) 
which makes use of the height of shower maximum to improve the core distance determination, 
and hence improve energy resolution to $<15$\%.
Both datasets \S1 and \S2 yielded consistent power-law fits with a hard photon index. For all data, 
a pure power law explains well the 
energy spectrum, showing no indication for a cut-off when fitting also a combined power law + exponential cutoff term
$\exp(-E/E_c)$.
We estimated, nevertheless, lower limits (99\% c.l.) to the cut-off energy for a range of (fixed) power law indices.
Cut-off lower limits of $E_c\sim3.6$, 4.2 and 4.6 TeV result when fixing the power index at values $\gamma=$1.7, 1.9, and 2.2, 
respectively. 

In estimating all flux values we have assumed a Gaussian source profile according to the 
estimated source size since the non-pointlike nature of the source is confirmed. Note that previously published
integral fluxes for combined data (Rowell et al. \cite{Rowell:2}, Horns et al. \cite{Horns:1}) assumed a 
pointlike source. 
It is also apparent that the event
rate for dataset \S2 is $\sim$80\% that of dataset \S1, and that the integral fluxes (for $E>1$ TeV) derived 
differ by about
40\%. However the statistical errors on the integral flux, which are dominated by contributions near 1 TeV,
suggests this difference is not significant ($\sim1\sigma$).  

We conclude therefore that TeV~J2032+4130 exhibited a constant flux from 1999 to 2002.
Note that the integral (and differential) fluxes are corrected for changes in the IACT-System response over time
(see P\"uhlhofer et al. \cite{Puhl:1} for details on the system performance over time). 
In this case, corrections up to the individual run level according to the CR background rate 
(a reliable, relative measure of the detector+atmospheric transmission) have been applied.

We also include in Table~\ref{tab:othersources}, 99\% upper limits for a number of other interesting sources in the FoV. 
These include
the two GeV sources, their possible associated X-ray counterparts (as indicated by Roberts et al. \cite{Roberts:1}), and
also Cygnus~X-3. 

\begin{table}[h]
\caption{Numerical summary for TeV~J2032+4130.
     (a) Centre of Gravity (CoG) and extension $\sigma_{\rm src}$
     (std. dev. of a 2D Gaussian); (b) Event summary. The values $s$ and $b$ are
     event numbers for the $\gamma$-ray-like and background (from the Template and Displaced models, see text) 
     respectively, 
     and $s-\alpha b$ is the excess using a normalisation $\alpha$. $S$ denotes the
     excess significance using Eq. 17 of Li \& Ma (\cite{Li:1}); 
     (c) Integral events after spectral cuts.}
\label{tab:numbers}

    {\centering \small
     \begin{tabular}{l}
     \fbox{\bf \boldmath (a) CoG \& Extension ($\epsilon\leq0.12^\circ$)}\\[2mm]
     \end{tabular}
     \begin{tabular}{lll}
     \multicolumn{3}{c}{---- All Data (278.3 h) ----}\\
      RA $\alpha_{2000}$:   &  20$^{\rm hr}$ 31$^{\rm m}$ 57.0$^{\rm s}$ & $\pm 6.2^{\rm s}_{\rm stat}$ $\pm 13.7^{\rm s}_{\rm sys}$ \\
      Dec $\delta_{2000}$:  &  41$^\circ$   \,\,29$^\prime$  \,\,56.8$^{\prime\prime}$ & $\pm 1.1^\prime_{\rm stat}$ $\pm 1.0^\prime_{\rm sys}$ \\
      $\sigma_{src}$        &  6.2$^\prime$  & $\pm 1.2^\prime_{\rm stat}$ $\pm 0.9^\prime_{\rm sys}$\\
     \end{tabular}\\[5mm]
     \begin{tabular}{l}\\
     \fbox{\bf \boldmath (b) Tight cuts: $\theta<0.12^\circ$, $\bar{w}<1.1$, $n_{\rm tel} \geq 3$} \\[2mm]
     \end{tabular}
     \begin{tabular}{lccccc} 
     Background & \\
     Model          &  $s$ & $b$   & $\alpha$ & $s-\alpha\,b$ & $S$ \\ \hline 
     \multicolumn{6}{c}{---- 1999 to 2001 Dataset \S1 (120.5 h) ----} \\
     Template       &   529 &  2432 & 0.168    &  123      & {\bf +5.3} \\        
     Displaced      &   529 &  6982 & 0.059    &  119      & {\bf +5.4} \\ 
     \multicolumn{6}{c}{---- 2002 Dataset \S2  (157.8 h) ----}\\
     Template       &   716 &  3494 & 0.168    &  129      & {\bf +4.8} \\        
     Displaced      &   716 &  8510 & 0.070    &  125      & {\bf +4.8} \\ 
     \multicolumn{6}{c}{---- All Data (278.3 h) ----}\\
     Template       &  1245 & 5926  & 0.168    &  252      & {\bf +7.1} \\        
     Displaced      &  1245 & 15492 & 0.065    &  243      & {\bf +7.1} \\ \hline 
   \end{tabular} \\[5mm]
   \begin{tabular}{l} \\
     \fbox{\bf (c) \boldmath Spectral Cuts: Tight Cuts + core$\leq 200$m}\\[2mm]
     Energy estimation method: See Hofmann et al. (\cite{Hofmann:3}) \\[2mm]
   \end{tabular}
   \begin{tabular}{lccccc}
     Background & \\
     Model   &  $s$ & $b$   & $\alpha$ & $s-\alpha\,b$ & $S$   \\ \hline
     \multicolumn{6}{c}{---- 1999 to 2001 Dataset \S1(120.5 h) ----}\\
     Displaced   &  421 & 2120  & 0.143   &   118        & {\bf +6.0} \\ 
     \multicolumn{6}{c}{---- 2002 Dataset \S2(157.8 h) ----}\\
     Displaced   &  552 & 2999  & 0.143   &   124        & {\bf +5.3} \\
     \multicolumn{6}{c}{---- All Data (278.3 h) ----}\\
     Displaced   &  973 & 5119  & 0.143   &   242        & {\bf +7.9} \\ \hline 
     \end{tabular}\\
   }
\end{table}

\begin{table}[h]
\caption{Numerical summary for TeV~J2032+4130 continued.. (d) differential fluxes and events after tight 
    spectral cuts; (e) Fitted power law; (f) Integral flux.}
\label{tab:numbers2}

    {\centering \small

     \begin{tabular}{c} 
     \fbox{\bf (d)  \boldmath Spectral Cuts: Differential points}\\[2mm]
     \end{tabular}
     \begin{tabular}{cccccc}
        Energy  & Flux $(E)^a$ & Flux$^a$   & $s$ & $b$ & $S^b$ \\ 
       $E$(TeV) &              & Error$(E)$ &     &     &       \\ \hline 
      \multicolumn{6}{c}{----  1999 to 2001 Dataset \S1(120.5 h) ----}\\
      1.05 & 1.70  & 1.24  & 130 & 678  & {\bf +3.0} \\
      1.82 & 1.94  & 0.91  & 107 & 531  & {\bf +3.1} \\
      3.16 & 0.37  & 0.27  & 57  & 323  & {\bf +1.4} \\
      5.50 & 0.30  & 0.11  & 33  & 117  & {\bf +3.2} \\
      9.55 & 0.09  & 0.04  & 13  &  34  & {\bf +2.8} \\
      \multicolumn{6}{c}{---- 2002 Dataset \S2(157.8 h) ----}\\
      1.05 & 19.90 & 16.57 & 153 & 885  & {\bf +2.1} \\
      1.82 & 1.85  & 1.56  & 152 & 913  & {\bf +1.7} \\ 
      3.16 & 0.85  & 0.28  & 107 & 513  & {\bf +3.4} \\ 
      5.50 & 0.28  & 0.10  & 44  & 174  & {\bf +3.2} \\ 
      9.55 & 0.07  & 0.04  & 17  & 51   & {\bf +2.8} \\
      \multicolumn{6}{c}{---- All Data (278.3 h) ----}\\
      1.05 & 11.98 & 10.80 & 283 & 1563 & {\bf +3.6} \\  
      1.82 & 1.89  & 0.97  & 259 & 1444 & {\bf +3.3} \\
      3.16 & 0.64  & 0.20  & 164 & 836  & {\bf +3.6} \\
      5.50 & 0.29  & 0.07  & 77  & 291  & {\bf +4.5} \\
      9.55 & 0.08  & 0.03  & 30  & 85   & {\bf +3.9} \\ \hline
      \multicolumn{6}{l}{\scriptsize a: Flux and Errors in units $\times 10^{-13}$ ph cm$^{-2}$s$^{-1}$ TeV$^{-1}$}\\
      \multicolumn{6}{l}{\scriptsize b: Significance from Li\&Ma (\cite{Li:1}) Eq. 17}\\
      \multicolumn{6}{l}{\scriptsize \hspace{3mm} using $s$, $b$ and a normalisation $\alpha$=0.143}\\
    \end{tabular}\\[5mm]
    \begin{tabular}{c} \\
     \fbox{\bf (e) Fitted Spectrum: Pure Power-Law}\\[2mm]
     \end{tabular}
     \begin{tabular}{ccl}\hline
      $dN/dE$ & = & $N\,\,(E/1\,{\rm TeV})^{-\gamma} \,\,\,\,{\rm ph\,\,cm^{-2}\,s^{-1}\,TeV^{-1}}$\\
      \multicolumn{3}{c}{----  1999 to 2001 Dataset \S1(120.5 h) ----}\\
      $N$     & = & $4.1\,\, (\pm2.1_{\rm stat} \pm1.3_{\rm sys})\, \times 10^{-13}$   \\
      $\gamma$& = & $1.7\,\, (\pm0.3_{\rm stat} \pm0.3_{\rm sys})$ \\ 
      \multicolumn{3}{c}{---- 2002 Dataset \S2(157.8 h) ----}\\
      $N$     & = & $9.3\,\, (\pm2.9_{\rm stat} \pm1.4_{\rm sys})\, \times 10^{-13}$   \\
      $\gamma$& = & $2.1\,\, (\pm0.2_{\rm stat} \pm0.3_{\rm sys})$ \\ 
      \multicolumn{3}{c}{---- All Data (278.3 h) ----} \\
      $N$     & = & $6.2\,\, (\pm1.5_{\rm stat} \pm1.3_{\rm sys})\, \times 10^{-13}$   \\
      $\gamma$& = & $1.9\,\, (\pm0.1_{\rm stat} \pm0.3_{\rm sys})$ \\ \hline
     \end{tabular}\\[5mm]
    \begin{tabular}{lcl} \\
     \fbox{\bf \boldmath (f) Integral Flux$^a$ (E$>$1 TeV)}\\[2mm]
     \end{tabular}
     \begin{tabular}{lcl}\hline
      \multicolumn{3}{c}{----  1999 to 2001 Dataset \S1(120.5 h) ----}\\
      $F (E>1\, {\rm TeV})$ & = & 5.86 $(\pm3.91_{\rm stat})$ \\
      \multicolumn{3}{c}{---- 2002 Dataset \S2(157.8 h) ----}\\
      $F (E>1\, {\rm TeV})$ & = & 8.45 $(\pm3.05_{\rm stat})$ \\
      \multicolumn{3}{c}{---- All Data (278.3 h) ----} \\
      $F (E>1\, {\rm TeV})$ & = & 6.89 $(\pm1.83_{\rm stat})$ \\ \hline
      \multicolumn{3}{l}{\scriptsize a: Flux and Errors in units $\times 10^{-13}$ ph cm$^{-2}$s$^{-1}$}\\

     \end{tabular}\\
   }
\end{table}

\begin{table}[h]
\caption{Summary of point-like flux upper limits for other sources in the FoV. Positions for 
         the various X-ray sources are taken from Roberts et al. (\cite{Roberts:1}). An energy threshold
         of $E_{\rm th}>0.7$ TeV is estimated based on the average zenith angle $\bar{z}$ of observations: 
              $E_{\rm th}=0.5\cos(\bar{z})^{-2.5}$. The template background model been used ($\alpha=0.168$), 
            with $s$, $b$ and $S$ defined as in table~\ref{tab:numbers}.}
\label{tab:othersources}
    {\centering \small
     \begin{tabular}{l}\\
     \fbox{\bf \boldmath Other Sources: (tight cuts) $\theta<0.12^\circ$, $\bar{w}<1.1$, $n_{\rm tel} \geq 3$} \\[2mm]
     \end{tabular}\\

     \begin{center}
     \begin{tabular}{lccccc} 
     Source       &  $s$ & $b$   & $s-\alpha\,b$ & $S$ & $^a\phi^{99\%}$ \\ \hline
     GeVJ2026+4124         & 541  & 3429 & $-$35 & \bf $-$1.3 & \bf 0.29 \\ 
     AXJ2027.6+4116        & 771  & 4497 & 16    & \bf +0.6   & \bf 0.42 \\ 
     GeVJ2035+4214         & 784  & 4457 & 35    & \bf +1.2   & \bf 0.45 \\ 
     AXJ2036.0+4218(Src1)  & 620  & 4005 & $-53$ & \bf $-$1.8 & \bf 0.22 \\ 
     AXJ2035.4+4222(Src2)  & 663  & 4083 & $-$23 & \bf $-$0.7 & \bf 0.27 \\ 
     AXJ2035.9+4229(Src3)  & 584  & 3626 & $-$25 & \bf $-$0.9 & \bf 0.28 \\ 
     Cygnus X-3            & 857  & 5613 & $-86$ & \bf $-$2.6 & \bf 0.17 \\ \hline  
     \multicolumn{6}{l}{\scriptsize a. $\phi^{99\%}_{ph} = $99\% upper limit $E_{\rm th}>0.7$ TeV [$\times 10^{-12}$ ph 
      cm$^{-2}$s$^{-1}$]}\\
   \end{tabular} 
   \end{center}
   }

\end{table}

\begin{figure*}
{\begin{center}
 \includegraphics[width=0.85\linewidth]{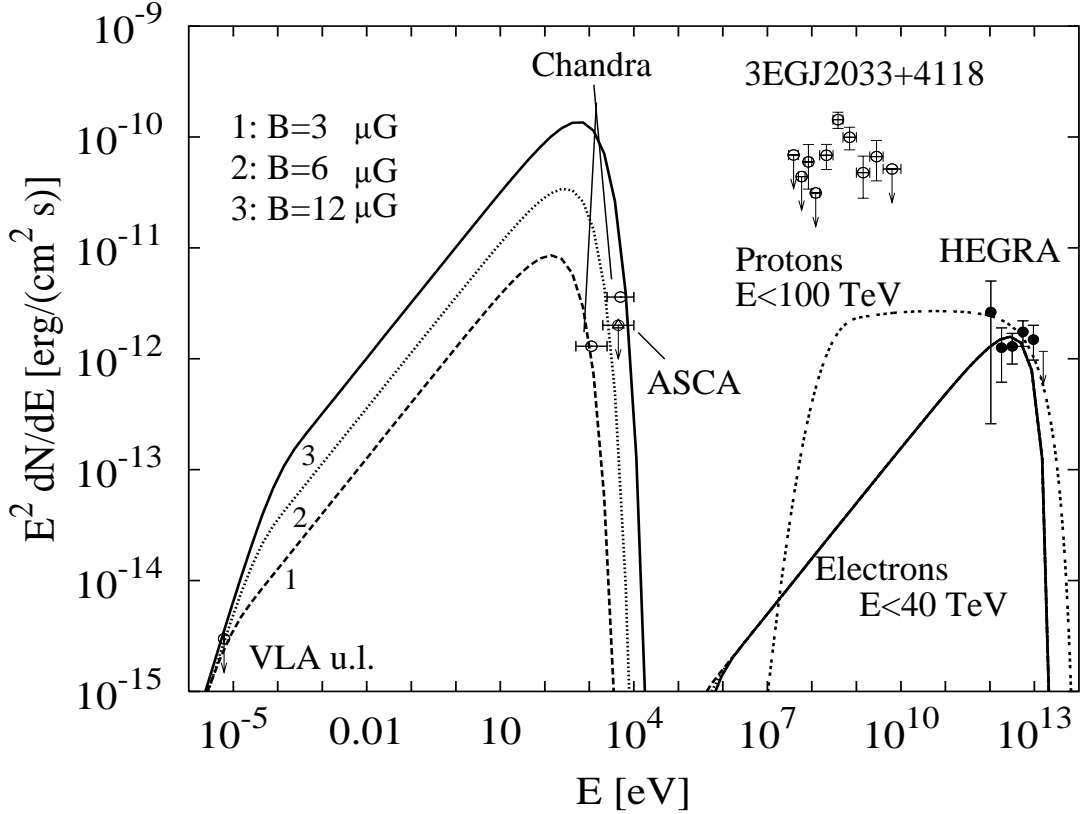}
 \end{center}
}
    \caption{
     Spectrum of TeV~J2032+4130 (this work - HEGRA) compared with
     purely hadronic (Protons E$<100$ TeV) and leptonic (Electrons E$<40$ TeV) models. 
     Upper limits, constraining the synchrotron emission (leptonic models), are from the VLA and 
     {\em Chandra} (Butt et al. \cite{Butt:1})  and ASCA (Aharonian et al. \cite{Aharonian:1}).
     In the model a minimum electron energy $\gamma_{\rm min}\sim 10^4$ is chosen to meet the VLA
     upper limit. 
     EGRET data points are from the 3rd EGRET catalogue (Hartman et al. \cite{Hartman:1}).} 
     \label{fig:spectrum} 
\end{figure*}

\section{Discussion and Conclusion}
Possible origins of TeV~J2032+4130 have been considered in Aharonian et al. (\cite{Aharonian:1}), 
Butt et al. (\cite{Butt:1}), Mukherjee et al. (\cite{Mukherjee:1}) and Bednarek (\cite{Bednarek:1}).
One interpretation involves association with the stellar
winds of member stars in Cygnus~OB2, individually or collectively, which provide conditions conducive to
strong and stable shock formation for particle acceleration.
Another scenario involves particle acceleration at a
termination shock, which is expected at the boundary where a relativistic jet meets the interstellar
medium. TeV~J2032+4130 in fact aligns well within the northern error cone of the bi-lobal jet of  
Cygnus~X-3  discussed by Mart\'{\i} et al. (\cite{Marti:1,Marti:2}).
The existence of TeV emission clearly suggests that particles are accelerated to at least multi-TeV energies. 
Taken at face value the different flux levels (a maximum factor $\sim$20 difference) claimed by the Crimean
(1.7 Crab $E>1$ TeV), HEGRA (0.05 Crab $E>1$ TeV) 
and Whipple (0.12 Crab $E>0.6$ TeV) groups over a period of a decade, would suggest episodic emission from TeV~J2032+4130.
Explaining {\bf also} the extended nature of TeV~J2032+4130 as seen by HEGRA requires consideration of issues such as the 
particle acceleration site and its distance to that of the TeV $\gamma$-ray emission, particle diffusion and source age. 
Such issues would also not in general rule out an extended {\em and} episodic source. Moreover the TeV~J2032+4130 emission 
at any given time may be a superposition of more than one component (eg. variable compact in addition to 
weak, steady, extended emission).

For illustrative purposes we have matched the spectral energy distribution of TeV~J2032+4130 with simplistic 
leptonic and hadronic models (Fig.~\ref{fig:spectrum}). We assume that the TeV emission arises from
a single pure population of either non-thermal hadronic or electronic parent particles. We do not consider here the
conditions under which particles are accelerated or how they lose energy. Under the hadronic scenario 
the $\pi^\circ$-decay prediction matches well the TeV flux using a parent proton power law 
spectrum of index $-$2.0 with a sharp limit at 100 TeV.
The neighbouring EGRET source 3EG~J2033+4118 (possibly not related to TeV~J2032+4130) should be considered here
as upper limits on the potential GeV flux of TeV~J2032+4130. 
Associated synchrotron X-ray emission would also be expected from tertiary electrons 
($\pi^\pm \ldots \rightarrow \mu^\pm  \ldots \rightarrow e^\pm$), not modeled at present,
which represent an absolute lower limit on any synchrotron emission visible
assuming a pure electronic scenario. TeV data are matched well by an 
inverse-Compton spectrum (from electrons up-scattering the cosmic microwave background) arising from
an electron spectrum with power law index $\sim\, -2.0$ and a sharp limit at 40 TeV. The
predicted synchrotron emission then follows as a function of local magnetic field, constrained by the available upper
limits at radio and X-ray energies. The most conservative synchrotron
prediction arises for $B_0=3 \mu$G, the lowest such B-field expected
in the Galactic disk. In fact, much higher B-fields ($B_0>10\mu$G) are generally expected in such regions 
containing young/massive stars with high mass losses and colliding winds (e.g. Eichler et al. \cite{Eichler:1}).
X-ray results from ASCA (Aharonian et al. \cite{Aharonian:1}) provide constraining upper limits, as do results from 
{\em Chandra} (Butt et al. \cite{Butt:1}).
Deeper observations by XMM and {\em Chandra} will no doubt provide further constraints on the leptonic component. Future
$\gamma$-ray observations by H.E.S.S., VERITAS and also MAGIC-II will be vital in better determining 
the spectrum over the 
$\sim$50 GeV to $>$10 TeV regime. Knowledge of the spectral behaviour for $E>$10 TeV will convey 
important information on the maximum particle energies and their type. Such information can come from low elevation
observations by H.E.S.S. for which very large collecting areas are achieved at TeV energies.
Energy-resolved morphology studies can also be performed, allowing conclusions on the diffusion 
properties of the accelerated particles. 
Overall, TeV~J2032+4130 is the only TeV source so far without an
obvious multiwavelength counterpart, and is likely the first-discovered galactic TeV source which is extended in nature. 

The serendipitous detection of such a weak ($\sim0.05$ Crab), marginally extended source with a hard spectrum 
over a long observation time illustrates the power of the stereoscopic technique as was employed 
by the HEGRA IACT-System. With a sensitivity 
at least a factor of 5 better, the next generation instruments will find such sources detectable in under 
10 hours, or even less for sources with steeper energy spectra.

Finally, we also obtained upper limits (for a steady TeV flux) from a number of other source positions in the 
FoV (Fig.~\ref{fig:skymap}, 
Table~\ref{tab:othersources}), including the famous microquasar Cygnus~X-3 for which an upper limit 
($E>0.7$ TeV) 1.7$\times$10$^{-13}$ ph cm$^{-2}$s$^{-1}$ for steady emission is set.

\begin{acknowledgements}
The support of the German Ministry for Research and
Technology BMBF and of the Spanish Research Council CICYT is gratefully
acknowledged.
We thank the Instituto de Astrof\'{\i}sica de Canarias
for the use of the site and for supplying excellent working conditions at
La Palma. We gratefully acknowledge the technical support staff of the
Heidelberg, Kiel, Munich and Yerevan Institutes. GPR acknowledges receipt of a von Humboldt fellowship.
\end{acknowledgements}

\end{document}